\input harvmac

\def\p{\partial}
\def\ap{\alpha'}
\def\half{{1\over 2}}

\def\O{{\cal O}}
\def\e{{\bf e}}
\def\x{{\bf x}}
\def\y{{\bf y}}
\def\k{{\bf k}}

\Title{EFI-99-13}{\vbox{\centerline{Energy-Momentum Conservation and
}
\vskip15pt
\centerline{Holographic S-Matrix
}}}
\vskip20pt

\centerline{Miao Li}
\centerline{\it Enrico Fermi Institute}
\centerline{\it University of Chicago}
\centerline{\it 5640 Ellis Avenue, Chicago, IL 60637} 
\centerline{\tt mli@theory.uchicago.edu}

\bigskip

We investigate the consequence of the energy-momentum conservation
law for the holographic S-matrix from AdS/CFT correspondence.
It is shown that the conservation law is not a natural consequence
of conformal invariance in the large N limit. We predict a new
singularity for the four point correlation function of a marginal
operator. Only the two point scattering amplitude is explicitly
calculated, and the result agrees with what is expected.
 
\Date{April 1999}

\nref\holo{G. 't Hooft, "Dimensional reduction in quantum gravity",
hep-th/9310026; L. Susskind, "The world as a hologram", 
hep-th/9409089.}
\nref\joe{J. Polchinski, `` S-Matrices from AdS Spacetime'',  hep-th/9901076.}
\nref\ls{L. Susskind, ``  Holography in the Flat Space Limit'',  
hep-th/9901079.}
\nref\other{V. Balasubramanian, S. B. Giddings and A. Lawrence,
``What Do CFTs Tell Us About Anti-de Sitter Spacetimes'', hep-th/9902052;
S. B. Giddings, ``The boundary S-matrix and the AdS to CFT dictionary'',
hep-th/9903048 .}
\nref\pst{J. Polchinski, L. Susskind and N. Toumbas, ``
Negative Energy, Superluminosity and Holography'', hep-th/9903228 .}
\nref\mald{J. M. Maldacena, ``The Large N Limit of Superconformal
Field Theories and Supergravity", hep-th/9711200.}
\nref\ew{E. Witten, ``Anti De Sitter Space And Holography'', hep-th/9802150.}
\nref\gkp{S. S. Gubser, I. R. Klebanov and A. M. Polyakov,
``Gauge Theory Correlators from Non-Critical String Theory'', 
hep-th/9802109.}
\nref\bkl{V. Balasubramanian, P. Kraus and A. Lawrence,
``Bulk vs. Boundary Dynamics in Anti-de Sitter Spacetime",
hep-th/9805171.}

\newsec{Introduction and Intuitive Observation}

Although there exist several nonperturbative formulations of M theory
and string theory in a flat spacetime, it has been hard to do a
quantitative calculation in the right regime where the formulation is
supposed to be valid. This difficulty has to do with the unusual physics
which is required of a holographic theory \holo. The most recent 
proposal is an explicit ansatz for the S-matrix in the flat space limit
using a convolution of conformal correlators in the boundary
conformal field theory \refs{\joe, \ls}, for a related discussion,
see \other. A few puzzling aspects
are already pointed out within this context in \pst.

The ansatz of \refs{\joe, \ls} involves taking a peculiar high energy
limit along with a large N limit. For instance, the type IIB
string theory defined on $R^{10}$ is encoded in the super Yang-Mills
theory on $S^3\times R$ in the large N limit. Of course the whole
theory may not exist in the large N limit when the Yang-Mills
coupling constant is held fixed. The conjecture of \refs{\joe,\ls}
rather asserts that a certain subsector must exist. If one possesses
infinite calculational power to calculate all the relevant correlation
functions in SYM, the IIB string S-matrix can be constructed
nonperturbatively.

Of course for the time-being we are not yet that powerful. On the
contrary, the Maldacena conjecture \mald\ has been used to make 
predictions about the SYM theory.  Although we do not know much 
about the nonperturbative S-matrix, a few principles are certainly
applicable. Traditionally, very general principles such as Lorentz
invariance, unitarity and analyticity  constitute strong constraints
on the S-matrix. We expect that these constraints transform into the
ones on a subset of correlation functions in SYM via the holographic
S-matrix ansatz. Our purpose in this paper is to
point out that the simplest consequence of Lorentz invariance,
the energy-momentum conservation law, is not a bona fide
consequence of conformal invariance. Notice that the isometry
of the anti-de Sitter space is the conformal group. Taking the
large radius limit, the conformal group contracts to the Poincare
group. Rather surprisingly, already
at the level of the 4 point amplitude, implementation of
energy-momentum conservation requires the existence of a new type of
singularity in the 4 point correlation functions in the large N
limit. This singularity, to our knowledge, is not dictated by
conformal symmetry.

In order to extract information about physical process happening
in the center of the anti-de Sitter space, well-focused wave
packets must be prepared. A precise ansatz for an incoming particle
or an outgoing particle is given in \joe. We will focus on massless
particles, for we will work with the type IIB string theory, and the
only stable states are those of supergraviton. Denote the creation
operator of an incoming particle by $\alpha_{\omega \e -}$, where
$\omega =RE$ is the dimensionless energy, $R$ is the radius of
$AdS_5$, and $\e$ is a unit 4 vector. This particle carries a momentum 
$\omega \e$ tangent to $AdS_5$ in its center. The state is smeared
over $S^5$, or it carries a zero momentum in the internal space.
Similarly, denote the annihilation operator of an outgoing particle
by $\alpha_{\omega \e +}$. For a scalar particle, the ansatz of
\joe\ is
\eqn\ans{\eqalign{\alpha_{\omega \e -}&=\omega^{-3/2}
\int dt d\Omega \exp\{-{\omega\over 2}[(t+\pi/2)^2+|\x +\e |^2]
-i\omega (t+\pi/2)\}\O(t, \x),\cr
\alpha_{\omega \e +}&=\omega^{-3/2}
\int dt d\Omega \exp\{-{\omega\over 2}[(t-\pi/2)^2+|\x -\e |^2]
+i\omega (t-\pi/2)\}\O(t, \x),}}
where $\O$ is the appropriate operator corresponding to
the scalar field \refs{\ew, \gkp, \bkl}. For the dilaton, it is proportional
to $\tr F^2$, for the axion, it is proportional to
$\tr F\wedge F$. We assumed that $\O$ is properly normalized, so
some $\omega$ independent numerical factors in \joe\ were dropped.
The integral $\int d\Omega$ is over the unit $S^3$ which is
parameterized by $\x$.

The above ansatz clearly indicates that the incoming (outgoing)
particle originates (ends up) at time $-\pi/2$ ($\pi/2$). In the
large $R$ limit, the proper time goes to $\pm \infty$.
The Gaussian factor helps to focus the beam in the direction $\e$. 
A S-matrix element is given by
\eqn\smt{S=\lim _{N\rightarrow\infty}\Phi^{-1}
\langle \prod_i\alpha_{\omega \e_i-}\prod_j  \alpha_{\omega \e_j +}
\rangle,}
where $\Phi$ is a normalization factor. For a fixed string coupling
constant, due to the relation $R^4=4\pi g_s N\ap^2 $, the large $R$ limit
is achieved by taking the large N limit.

In a conformal field theory, both the two point functions and
three point functions are fixed up to a numerical coefficient
by conformal symmetry. One would expect that the calculation
of the two point amplitudes and three point amplitudes using
eqs.\ans, \smt\ is a simple matter. Actually, as shown in the
next section, an exact form of three amplitude can be obtained only
after some tedious calculation. In this section we will be
content with a qualitative examination of these amplitudes.

The geometry $S^3\times R$ is conformal to $R^4$, so correlation
functions on $S^3\times R$ can be obtained from those on $R^4$
using the conformal transformation. For instance, the Euclidean
distance between the two points $x$ and $y$, $r^2=|x-y|^2$
is mapped to 
\eqn\mapp{\exp (\tau_x+\tau_y)\left(\cosh (\tau_x-\tau_y)-\x\cdot \y
\right),}
where we parametrize $R^4$ by the radial coordinates $r=\exp\tau$
and the unit sphere $S^3$. $\x$ and $\y$ are unit 4 vectors.
Now $\tau$ and $\x$ parametrize $S^3\times R$. 

In a correlation function, the extra factors such as $\exp(\tau_1
+\tau_2)$ are removed by a conformal factor. If $F(r^2_{ij})$
is a correlation function on $R^4$, then the corresponding
correlation function on $S^3\times R $ is obtained by simply
replacing $r^2_{ij}$ with $\cosh (\tau_i-\tau_j)-\x_i\cdot\x_j$.
To obtain the correlation function on a Minkowskian $S^3\times R$,
we wick-rotate $\tau$ to $it$, and add a term $i\epsilon$:
\eqn\pmin{r_{ij}^2=\cos (t_i-t_j)-\x_i\cdot \x_j +i\epsilon,}
where the $i\epsilon$ prescription is introduced to ensure causality
in the boundary conformal theory.
To see this, assume $t_i-t_j$  small, and the angle $\phi_{ij}$
between $\x_i$ and $\x_j$ small, we obtain
\eqn\cas{-\half (t_i-t_j)^2+\half \phi_{ij}^2 +i\epsilon,}
we see that if one uses $r_{ij}^{-2}$ as the propagator, the signal
will propagate along the future light-cone for a positive energy
mode of the form $\exp(-i\omega t)$.

The scaling dimension of an operator corresponding to a massless scalar 
field is $\Delta =4$. The two point function is therefore
\eqn\tpf{\langle \O (t_1,\x_1)\O (t_2,\x_2)\rangle =
(\cos (t_1-t_2)-\x_1\cdot \x_2+i\epsilon )^{-4}}
up to a normalization constant.

The two point scattering amplitude is obtained using \smt. Without
taking the large N and high energy limit, there is no energy-momentum
conservation, since as shown in the appendix of \joe, the incoming
an well as outgoing waves have finite width in both energy and
momentum. The width of $\omega$ is proportional to $\sqrt{\omega}$.
Using $\omega =R E$, the width of $E$ is proportional to
$\sqrt{E/R}$ and goes to zero in the large R limit. So energy-momentum
conservation has to be recovered in this limit.
We shall show in the next section that the energy conservation is
always guaranteed in this limit.

However, the momentum conservation
rather imposes strong constraints on the behavior of conformal correlators
in the large N limit. As we will see in the next section, the convolution
of \smt\ using ansatz \ans\ is rather subtle. To obtain the exact numerical
answer, one cannot simply replace the Gaussian distribution of \ans\
by a simpler one, say a delta function. To see the momentum conservation,
though, we will do this in this section.

In the case of two point amplitude, replacing the Gaussian wave
packets by delta functions, we obtain
\eqn\simplit{(1-\e_1\cdot \e_2)^{-4}.}
Together with a factor depending on $\omega_i$, we will have a null
result if $\e_1\cdot\e_2\ne 1$. The above expression is singular
if $\e_1\cdot \e_2=1$ or equivalently $\e_1=\e_2$. Thus we hope that a more
careful calculation will result a delta function $\delta^3(\e_1
-\e_2)$ for the two point amplitude. Here we define the delta function
by
$$\int \delta^3(\e_1-\e_2)d\Omega_2=1.$$
Together with the energy conservation $\omega_1=\omega_2$, or 
$E_1=E_2$, this implies the momentum conservation $E_1\e_1=E_2\e_2$.
The conservation is due to the fact that the two point conformal
correlation function is singular if one point sits on the future
light-cone of the other point.

The three point correlation function of operator $\O$ is fixed by
conformal invariance. Again up to a constant, it is
\eqn\tpf{\eqalign{&\langle\O (t_1,\x_1)\O (t_2,\x_2)\O (t_3,\x_3)\rangle
=(\cos (t_1-t_2)-\x_1\cdot \x_2+i\epsilon)^{-2}\cr
&(\cos (t_2-t_3)
-\x_2\cdot\x_3+i\epsilon )^{-2} (\cos (t_1-t_3)-\x_1\cdot\x_3
+i\epsilon)^{-2}.}}

Consider the case where two states are incoming, one is outgoing.
The Gaussian wave packets force $t_{1,2}$ to center around
$-\pi/2$, and $t_3$ to center around $\pi/2$.
If we simply replace the Gaussian wave packets by delta functions,
the above expression becomes
\eqn\tsin{(1-\e_1\cdot\e_2)^{-2}(1-\e_2\cdot\e_3)^{-2}(1-\e_1\cdot
\e_3)^{-2}.}
This function is singular whenever $\e_i=\e_j$ for $i\ne j$.
It is most singular when all $\e_i$ are equal. Thus we expect
that a more careful treatment will lead to delta functions
forcing all $\e_i$ to be equal. Indeed this is the consequence of
momentum conservation for a three point amplitude involving
only massless particles: That all momenta of outlegs must be collinear.
This is most easily seen in the following physical way. If, say,
the two incoming states are not collinear, then one can go to the
center of mass frame in which the end product of the scattering
can never be a single massless particle.

Mathematically, the above result, as in the two point amplitude case,
is a consequence of causality in the boundary theory: Whenever
two points are separated by a null geodesics, then the correlation
function becomes singular. For the three point function, it is
most singular when they lie on the same light-cone.

This raises a puzzle already at the level of the three point amplitudes.
Imagine that there are stable massive particles. In this case
the three point correlation is still given by a formula similar
to \tpf\ if one replaces $2$ by $(\Delta_i+\Delta_j-\Delta_k)/2$, 
if the particles
carry zero momentum in the internal space $S^5$. Again the correlation
function becomes singular when $\e_i$ and $\e_j$ are equal,
provided $\Delta_i+\Delta_j>\Delta_k$.
However this has nothing to do with the momentum conservation
for massive particles. It is therefore quite interesting to note
that there are no stable massive stringy states in the type IIB
theory. Thus this puzzle is avoided. A related feature is that
the conformal dimension of a stringy state is divergent in the large
N limit \refs{\ew,\gkp}, thus although the convolution \smt\ 
exists for finite N, its large N limit does not exist.

It is more interesting to see what happens to the four point
correlation function when the momentum conservation is imposed.
The four point correlation function of operator $\O$ of scaling
dimension 4 can not be fixed by conformal symmetry alone. Up
to a scaling factor, it is a function of two independent cross-ratios.
Define the cross-ratios
\eqn\crt{a={r^2_{12}r^2_{34}\over r^2_{13}r^2_{24}}, \quad
b={r^2_{12}r^2_{34}\over r^2_{23}r^2_{14}},}
the correlation function can be written as, for instance
\eqn\fpt{\eqalign{&\langle \O(t_1,\x_1)\dots \O (t_4,\x_4)
=F_4(r^2_{ij})\cr
&=\prod_{i<j\le 4}r_{ij}^{-8/3}f(a,b),}}
where $f$ is a undetermined function of $a$ and $b$. All $r^2_{ij}$
are as given in \pmin.

Unlike in some 2D conformal field theories, $f(a,b)$ is not
constrained in SYM as far as we know, since no other nontrivial
symmetries extending conformal invariance have been discovered
by far. The scaling factor in \fpt\ is singular whenever two
points are separated by a null geodesics. $f(a,b)$ can be singular
too in this case, since one of $a$ and $b$ vanishes or becomes
infinity. Unlike for the two and three point amplitudes, energy-momentum
conservation in general does not require two points being 
on the same light-cone. To see the general consequence, we
follow the same strategy as the above to replace $r_{ij}^2$
by their ``on-shell" value:
\eqn\osv{\eqalign{r_{12}^2&=1-\e_1\cdot\e_2=2\sin^2(\phi_{12}/2),
\quad
r^2_{34}=2\sin^2(\phi_{34}/2), \cr
r^2_{13}&=-1+\e_3\cdot\e_4=-2\sin^2(\phi_{13}/2),\quad
r^2_{24}=-2\sin^2(\phi_{24}/2),\cr
r^2_{23}&=-2\sin^2(\phi_{23}/2),\quad r^2_{14}=-2\sin^2(\phi_{14}/2),}}
where we assumed that particles 1 and 2 are incoming, and particles
3 and 4 are outgoing;
$\phi_{ij}$ is the angle between momentum $\k_i$ and momentum
$\k_j$.

Now use the energy conservation $E_1+E_2=E_3+E_4$ and the momentum
conservation law $E_1\e_1+E_2\e_2=E_3\e_3+E_4\e_4$, we derive
\eqn\mans{s=4E_1E_2\sin^2(\phi_{12}/2)=4E_3E_4\sin^2(\phi_{34}/2),}
where $s$ is one of the Mandelstam variables. Similarly
\eqn\mantu{\eqalign{t&=-4E_1E_3\sin^2(\phi_{13}/2)=
-4E_2E_4\sin^2(\phi_{24}/2),\cr
u&=-4E_1E_4\sin^2(\phi_{14}/2)=-4E_2E_3\sin^2(\phi_{23}/2).}}
We see that those ``on-shell" distances in \osv\ are simply related
to the Mandelstam variables. However, $r^2_{ij}$ also depends on
individual energies. It appears that the only way to eliminate
the dependence on energies is to
use the two cross-ratios
\eqn\crtio{a={s^2\over t^2},\quad b={s^2\over u^2}.}
Now since Mandelstam variables satisfy relation $s+t+u=0$, we obtain
\eqn\main{{1\over \sqrt{a}}+{1\over\sqrt{b}}=1.}
In other words, the above relation is a consequence and energy-momentum
conservation.

Our experience with the two and three point amplitudes tells us that
in order for the 4 point scattering amplitude to obey energy-momentum
conservation, the four point correlation function must be singular
when \main\ is satisfied. The correlation is also singular whenever
one of $r^2_{ij}$ vanishes.
As already explained, $r^2_{ij}=0$ has
nothing to do with energy-momentum conservation, the singularity
of $f(a,b)$ at $1/\sqrt{a}+1/\sqrt{b}=1$ must be severe
than the singularity of the correlation at $r^2_{ij}=0$.

The new singularity we observed above is not dictated by conformal
invariance at all. Also, this singularity is demanded only in
the large N limit, since for finite N there is no momentum conservation
in the anti-de Sitter space. As we will explain in detail in the
next section, $f(a,b)$ must be singular if
\eqn\amain{4ab=(ab-a-b)^2,}
and \main\ is only one of the four solutions 
\eqn\aamain{{1\over\sqrt{a}}\pm{1\over\sqrt{b}}=\pm 1}
to the above equation.

\newsec{Explicit Calculations}

We will see that an explicit calculation based on the ansatz of \joe\ is quite
difficult. We will be able to obtain a closed form for the two pint amplitude,
we will not be able to complete the calculation of the three point amplitude.
However, we will show that energy-momentum conservation is ensured in the large
R limit. We discuss the calculation of the four point amplitude. A lot of work
is left for future.

First, we want to simplify the calculation of a general amplitude a bit. 
Shifting $t\rightarrow t\mp \pi/2$, for an incoming particle or an 
outgoing one, one can always ignore the quadratic term in $t$ in ansatz
\ans, if one remembers that the integral over $t$ will always pick
up the most important contribution around $t\sim 0$. To see this, consider
the integral
\eqn\simp{\int dt e^{-{\omega\over 2}t^2+i\omega t}f(t,\dots),}
where the dots denote other variables. $f(t)$ is a periodic function
of $t$ of period $2\pi$. Express $f(t,\dots)$ as
$$f(t,\dots )=\int d\omega \tilde{f}(\omega,\dots) e^{-i\omega t},
$$
then
\eqn\fint{\int dt e^{-{\omega\over 2}t^2+i\omega t}f(t,\dots )
=\sqrt{{2\pi\over\omega}}\int d\omega' e^{-{(\omega-\omega')^2\over
\omega}}\tilde{f}(\omega',\dots ).}
The integral over $\omega'$ centers around $\omega$ with width
$\delta\omega'\sim\sqrt{\omega}$, compared with the principle value
$\omega$, this deviation tends to zero in the large $\omega$ limit.
Thus the above integral is approximately
\eqn\apro{2\pi \tilde{f}(\omega,\dots)=\int dt e^{i\omega t}
f(t,\dots ).}
Compared with this approximate value, the deviation is about
\eqn\dev{\delta\omega {d\tilde{f}(\omega,\dots )\over d\omega},}
so it can be ignored if 
$$\tilde{f}'(\omega,\dots )/\tilde{f}(\omega,\dots )\ll 1/\sqrt{\omega}.$$
This condition is generally satisfied.
Alternatively, as we already saw, the function $f(t,\dots)$ 
has infinitely many poles. Integral over $t$ will pick up these
poles, the Gaussian factor $\exp (-(\omega/2)t^2)$ helps to
suppress all these poles except the one closest to zero. In this case,
the other factor $\exp (i\omega t)$ is more important than the
Gaussian factor, since it oscillates fast around $t\sim
1/\sqrt{\omega}$.

To demonstrate the above result, let us see how the energy conservation
is derived in the large $\omega$ limit. For a general scattering
amplitude, we have
\eqn\gamp{\int\prod_idt_i e^{-{1\over 2}\sum_i\omega_i t_i^2
+i\sum_i a_i\omega_i t_i}f(t_i,\dots),}
where $a_i=1$ for an outgoing state, and it is $-1$ for an incoming
state. Use a new set of times: $t_1$, $t_i=\tau_i+t_1$, $i=2,\dots,
n$. The function $f(t_i,\dots)$ is a function of $\tau_i$ only,
since the correlation function in SYM is invariant under a time
translation. Performing the integral over $t_1$ first, we obtain the
integrand for $\tau_i$
\eqn\resl{\sqrt{2\pi\over \omega}e^{-{1\over 2\omega}(\sum a_i\omega_i)^2
-{1\over 2\omega}[\sum \omega_i\tau_i]^2-{1\over 2}\sum \omega_i\tau_i^2
+i\sum a_i\omega_i\tau_i}f(\tau_i,\dots ),}
where $\omega =\sum\omega_i$, and we omitted a term 
$$-{i\over\omega}(\sum a_i\omega_i)(\sum\omega_i\tau_i)$$
in the exponential. This is because the first exponential together
with the prefactor gives rise to a delta function in the large R
limit
\eqn\enc{{2\pi\over R}\delta(\sum a_i E_i),}
which is just the energy conservation law. This factor can be
obtained without including the Gaussian factors. The remaining Gaussian
factor in \resl\ is positive definite in $\tau_i$, and as we argued
before, can be ignored so long if we pick up poles closest to
zero in $\tau_i$.

The integral over variables on $S^3$ is much more complicated, and
since the Gaussian factor $\exp(-(\omega_i/2)|\x_i\pm \e_i|^2)$
is the only nontrivial factor in the convolution, one has to treat
the convolution carefully. In the following, we will examine
the two, three and four amplitudes separately. Before doing that,
let us remark that this Gaussian factor can be replaced by
\eqn\rga{\eqalign{&e^{-{\omega_i\over 2}|\x_i \pm\e_i|^2}=
\int {d^4k_i\over (2\pi\omega_i)^2}e^{-{k^2_i\over 2\omega_i}
+ik_i(\e_i\pm \x_i)}\cr
&=\int {d^4k_i\over (2\pi)^2}e^{-{k^2_i\over 2}+i\sqrt{\omega_i}
k_i\cdot (\e_i\pm\x_i)}.}}

\subsec{Two Point Amplitude}

As shown above, the integration over the ``center of times" results
in a factor
\eqn\rew{{2\pi\over R}\delta (E_1-E_2).}
The remaining part is
\eqn\remain{\Phi^{-1}N_2(RE_1)^{-3}\int {d^4k_1\over (2\pi)^2}{d^4k_2\over
(2\pi)^2}e^{-k_1^2-k_2^2/2 +i\sqrt{\omega_1}\sum k_i\cdot \e_i}
F_2(k_i),}
with
\eqn\newf{F_2(k_i)=\int dtd\Omega_1 d\Omega_2e^{-i\omega_1 t
-i\sqrt{\omega_1}\sum a_ik_i\cdot \x_i}(\cos t -\x_1\cdot \x_2 -i\epsilon)^{-4},}
where in eq.\remain\ we introduced a normalization factor $N_2$ depending
on a normalization of operator $\O$. 

The integral over $t$ in \newf\ can be performed first. It picks up
a pole at $t=\phi_{12}-i\epsilon$, where $\x_1\cdot\x_2=\cos\phi_{12}$.
Other poles are suppressed by a Gaussian factor we have omitted.
The leading contribution is
\eqn\ldc{F_2(k_i)={\pi\over 3}\omega_1^3\int d\Omega_1 
d\Omega_2 e^{-i\Omega \phi_{12}
-i\sqrt{\omega_1}\sum a_ik_i\cdot\x_i}{1\over \sin^4(\phi_{12}-i
\epsilon )},}
other terms are suppressed by powers of $\omega_1$.
Performing the integration over $\Omega_2$ first, we have
\eqn\nmo{F_2={8\pi^3\over 3}\omega_1^4\int d\Omega_1
e^{i\sqrt{\omega_1}(k_2-k_1)\cdot
\x_1}={32\pi^4\over 3}\omega_1^4 {\pi J_1(\sqrt{\omega_1} |k_1-k_2|)
\over \sqrt{\omega_1} |k_1-k_2|}.}
We will see momentarily that the fact that the above result is a function
of only $k_1-k_2$ ensures the momentum conservation in the large
R limit.

Plugging back the above result into \remain, ignoring the prefactor
in \remain\ for the moment, we have, after changing variables
$k_1-k_2=\kappa_2/2$, $k_1+k_2=\kappa_1/2$
\eqn\pint{{2^5\pi\over 3}\omega_1^4\int d^4\kappa_1
e^{-\kappa_1^2+i\sqrt{\omega_1}\kappa_1\cdot (\e_1-\e_2)}
\int {d^4\kappa_2\over 2\sqrt{\omega_1}|\kappa_2|}e^{-\kappa_2^2
+i\sqrt{\omega_1}\kappa_2\cdot (\e_1+\e_2)}J_1(2\sqrt{\omega_1}|\kappa_2|).}
The integral over $\kappa_1$ is separated from that over $\kappa_2$.
The first integral results in
$$\pi^2e^{-\omega_1 |\e_1-\e_2|^2}$$
which in the large $\omega$ limit tends to
\eqn\dta{\pi^2\left({\pi\over\omega_1}\right)^{3/2}\delta^3(
\e_1-\e_2).}
Due to the delta function, the second integral in \pint\ is simplified.
The integral over $\kappa_2$ can be separated into the radial
part and the angular part, and the latter can be easily performed.
In the end, we obtain
\eqn\endr{{2^8\pi^5\over 3}(\pi\omega_1)^{3/2}\delta^3(\e_1
-\e_2)\int dk k J_1^2(
2\sqrt{\omega_1}k)e^{-k^2}.}
In the large $\omega_1$ limit, the Bessel function $J_1$ can be replaced
by its asymptotic form, namely
$$\int_0^\infty dk k J^2_1(2\sqrt{\omega_1}k)e^{-k^2}
\rightarrow {1\over \pi\sqrt{\omega_1}}\int dk \cos^2(2\sqrt{\omega_1}
k)e^{-k^2},$$
and since for a large $\omega_1$, the cos factor can be replaced by
its average value $1/2$, the value of the above integral is 
$(1/4)(1/\sqrt{\pi\omega_1})$. We have checked this result by
using a formula for the integral in \endr\ involving the
Bessel function. Substitute this into \endr, we find
\eqn\midr{{(2\pi)^6\over 3}\omega_1\delta^3(\e_1-\e_2).}
Together with the prefactor in \remain\ and the delta function
in \rew, the end result is
\eqn\twp{\langle \alpha_{\omega_1\e_1-}\alpha_{\omega_2\e_2+}
\rangle =\Phi^{-1}N_2{(2\pi)^7\over 3R^3}E_1^{-2}
\delta (E_1-E_2)\delta^3(\e_1-\e_2).}
Using the identity
$$\delta^4(E_1\e_1-E_2\e_2)=E_1^{-3}\delta (E_1-E_2)\delta^3
(\e_1-\e_2)$$
it follows that
\eqn\ennd{\langle\alpha_{\omega_1\e_1-}\alpha_{\omega_2\e_2+}
\rangle =\Phi^{-1}N_2{(2\pi)^7\over 3R^3}E_1\delta^4
(E_1\e_1-E_2\e_2).}

Conservation of momentum for the two point amplitude also ensures
conservation of energy, so we have obtained the right delta
function, as expected by our intuitive argument in the last
section. It remains to check whether the kinetic factor is also
right. Note that the first normalization factor in \ennd\ depends
on the overlap of the two wave functions \joe, so it can depend
on the energy. The other normalization factor, $N_2$, does not
depend on energy, although it can be a function of $R$ and $g_s$.

The wave function of \joe\ takes the form
\eqn\tform{F(t,x)=e^{-i\omega(t-\e\cdot x)-{\omega\over 2}
(x_{\perp}^2+(t-\e\cdot x)^2)}}
near the center of AdS space.
The Gaussian factor can be ignored so long if the spacetime region
has a scale smaller than $1/\sqrt{\omega}$. The proper scale
is $R/\sqrt{\omega}\sim\sqrt{R}$. So in the large $R$ limit
the Gaussian factor can be ignored and we have a plan wave.
Since the creation operator is defined by
$$\alpha=\int dV\hat{\phi}\p_t F(t,x)+\dots,$$
so close to the center of the AdS space we have, roughly
\eqn\sjan{\hat{\phi}(t,x)=\int {d^4k\over E}\left(\alpha^+ (k)
e^{-iEt+ik\cdot x}+\alpha (k)e^{iEt-k\cdot x}\right),}
where all the coordinates are the proper ones, unlike the ones
in \tform. Thus we expect
that the scattering amplitude, up to a numerical factor, must be
\eqn\appx{\langle\alpha_{\omega_1\e_1-}\alpha_{\omega_2\e_2+}\rangle
=R^5E_1\delta^4(E_1\e_1-E_2\e_2),}
where the volume factor $R^5$ comes from the internal space $S^5$,
since the particles have zero momenta in the internal space. The
kinetic factor is precisely the same as in \ennd.
Therefore, it appears that the normalization $\Phi$ is order 1, and
$N_2\sim R^8\sim N^2g_s^2$.

\subsec{Three Point Amplitude}

Consider the three point amplitude with two incoming particles.
It is more convenient to integrate out $t_3$ first, with
$t_1=\tau_1+t_3$, $t_2=\tau_2+t_3$. As before a delta function
ensuring energy conservation results. The amplitude is
\eqn\tamp{A_3=N_3{2\pi\over R}\delta (E_1+E_2-E_3)\prod_i\omega_i^{-3/2}
\int \prod_i{d^4k_i\over (2\pi)^2}e^{-1/2\sum k_i^2+i\sum\sqrt{\omega_i}
k_i\e_i}F_3(k_i),}
with
\eqn\fthr{\eqalign{F_3(k_i)&=\int d\tau_1 d\tau_2\prod_id\Omega_i
e^{-i\omega_i\tau_i+i\sqrt{\omega_i}k_i\cdot\x_i}\cr
&(\cos (\tau_1-\tau_2)-\x_1\cdot\x_2+i\epsilon )^{-2}
(\cos \tau_1-\x_1\cdot\x_3-i\epsilon )^{-2}(\cos \tau_2-\x_2\cdot
\x_3-i\epsilon )^{-2},}}
where we have reflected $\x_3\rightarrow -\x_3$. Denote
$\x_i\cdot\x_j$ by $\cos\phi_{ij}$.

To perform the integral over $\tau_i$ first, we use the following
formula
\eqn\aux{(\cos\tau -\cos\phi \pm i\epsilon)^{-2}
=-{1\over\sin^2(\phi\pm i\epsilon)}\int^\infty_{-\infty}
d\omega |\omega|e^{i\omega\tau \pm i|\omega|(\phi\pm i\epsilon )},}
which is derived from
\eqn\auxi{(\cos \tau-\cos\phi\pm i\epsilon )^{-1}
=\mp {i\over \sin (\phi\pm i\epsilon )}\int d\omega 
e^{i\omega \tau \pm i|\omega |(\phi \pm i\epsilon )}}
by taking derivative with respect to $\phi$ once. With the Fourier
transform \aux, the integral over $\tau_i$ in \fthr\ is readily
performed, with the result
\eqn\redp{\eqalign{&F_3(k_i)=-\int\prod_i d\Omega_i
{\exp (i\sqrt{\omega_i}k_i\cdot \x_i)\over \sin^2(\phi_{12}+i
\epsilon)\sin^2(\phi_{13}-i\epsilon)\sin^2(\phi_{23}-i\epsilon)}
\cr
&\int d\omega |\omega (\omega_1-\omega )(\omega_2+\omega )|
e^{i|\omega |\phi_{12}-i|\omega_1-\omega |\phi_{13}
-i|\omega_2+\omega |\phi_{23|}}.}}

We do not know how to carry out the calculation of the above
integral. One thing is certain, though, that due to the singular
behavior of the integrand, the integral is peaked around
$\phi_{12}=\phi_{13}=\phi_{23}=0$. We thus expect that $F_3$ will
be a function of $\sum\sqrt{\omega_i}k_i$ only. We now argue that
this ensures the momentum conservation in the large R limit.
Introduce new vectors
\eqn\newv{\eqalign{l_1 &={1\over\sqrt{2\omega_3}}\sum \sqrt{\omega_i}
k_i,\cr
l_2 &=-\sqrt{\omega_2/\omega_3}k_1+\sqrt{\omega_1\omega_3}k_2 \cr
l_3 &=\sqrt{\omega_1/(2\omega_3)}k_1+\sqrt{\omega_2/(2\omega_3)}
k_2-1/\sqrt{2} k_3.}}
With the above relations, it is easy to see that $\sum k_i^2
=\sum l_i^2$, using the fact $\omega_3=\omega_1+\omega_2$.
Now $F_3$ is a function of $l_1$ only. We can perform
the integral in \tamp\ over $l_2$ and $l_3$ first. We note
in particular that in the exponential in \tamp, the $l_3$
dependent part is
\eqn\ltd{-{1\over 2} l_3^2 +i{l_3\over \sqrt{2\omega_3}}(\sum a_i
\omega_i \e_i).}
It is seen that the integral over $l_3$ results in a delta
function ensuring momentum conservation, in the large R limit.

After some calculation, we obtain
\eqn\endr{\eqalign{A_3&=N_3{2^{11/2}\pi^{5/2}E_3^2\over
E_1^3E_2^3}\delta (\sum a_iE_i)\delta^4(\sum a_i E_i\e_i)
\cr
&\int {d^4l_1\over (2\pi)^2}e^{-1/2l_1^2+i\sqrt{2\omega_1}l_1\cdot
\e_1}F_3(l_1).}}

\subsec{Four Point Amplitude}

We have much less to say about the four point amplitude. Although
we trust that the condition for energy-momentum conservation derived
in the previous section is necessary, we are not able to prove
that it is also sufficient. As in the two and three point amplitude
cases, we can always write the four point amplitude as
\eqn\famp{A_4=N_4\int\prod_{i=1}^4{d^4k_i\over (2\pi)^2}
e^{-1/2 \sum k_i^2 +i\sum\sqrt{\omega_i}k_i\cdot\e_i}F_4(k_i),}
where $F_4$ is given by a similar formula as \fthr. We omitted
a factor conserving energy. 

To see whether the momentum conservation is true in the large R limit,
as in the previous subsection, we introduce a set of new vectors
\eqn\sen{l_i=\sum_j\Omega_{ij}k_j,}
with $\{\Omega_{ij}\}$ being an orthogonal matrix. We can choose
$\Omega_{1i}=a_i\sqrt{\omega_i/\omega}$, where $\omega=\sum_i\omega_i$.
If in the large R limit, $F_4(l_i)$ is less dependent on $l_1$
than on other $l_i$, the integral over $l_1$ in \famp\ can be
performed first, thus resulting in a delta function associated with
the momentum conservation.

It can be shown that the condition that $1/\sqrt{a}+1/\sqrt{b}=1$
is a singularity of $f(a,b)$ is a necessary one, where
$f(a,b)$ is a function introduced in \fpt. It is far from
clear whether it is also the sufficient condition for conserving
momentum.

Finally, we want to show that if $1/\sqrt{a} +1/\sqrt{b}=1$ is a 
singularity of $f(a,b)$, then both $1/\sqrt{a}-1/\sqrt{b}=1$
and $1/\sqrt{a}-1\sqrt{b}=-1$ are also singularities of $f(a,b)$.
$f(a,b)$ is a symmetric function of $a$ and $b$. To see this,
we go to the Euclidean space. Exchange point 1 with point 2,
$a$ and $b$ are exchanged. Exchange point 2 and point 3
we are led to $a\rightarrow 1/a$, $b\rightarrow b/a$, thus
\eqn\ean{F(1/a, b/a)=F(a,b)=F(b,a).}
Now, $1/\sqrt{a}+1/\sqrt{b}=1$ is the same as $\sqrt{a}
-\sqrt{a/b}=1$. This means that $f(1/a,b/a)$ is singular when
this relation is satisfied. Renaming the variables, we 
conclude that $f(a,b)$ is singular when $1/\sqrt{a}-1/\sqrt{b}=1$.
Exchanging $a$ with $b$, we deduce that $f(a,b)$ is singular
when $1/\sqrt{a}-1/\sqrt{b}=-1$.

\newsec{Conclusion}

We have only scratched the surface of the problem of investigating
the consequence of Lorentz invariance for the holographic S-matrix and
associated correlation functions in the large N limit. For
instance, more constraints can be derived from the requirement
that $A_4$ is a function of only $s$ and $t$, apart from a kinetic
factor. Even more interesting, is the consequence of causality
in the flat space limit. We leave these problems for future
investigations.

Our main result in this paper is the identification of a new
singularity in the four point amplitude, in the large N
limit. This means that the
dominant contribution to the scattering amplitude comes from
around this ``saddle point". This reminds us the problem of 
sensitive initial conditions raised in \pst. It is observed
there that if the two beams aimed at the center of the AdS
are emitted from the boundary with a time difference greater 
than $1/R$, then the beams will miss each other. In the large 
N limit, this time
difference can be arbitrarily small. It appears that a kind
of sharp saddle point may help to understand this puzzle.
Presumably this is a consequence of locality in the bulk space.
It remains to see whether bulk locality together boundary
conformal invariance guarantee bulk Poincare invariance in
the large N limit.

\noindent {\bf Acknowledgments}
This work was supported by DOE grant DE-FG02-90ER-40560 and NSF grant
PHY 91-23780. 
We are grateful to R. Siebelink for collaboration during the initial
stage of this project. Useful discussions with P. Kraus have
helped me going through some difficult phase. A correspondence with
J. Polchinski is acknowledged. We are also grateful to P-M. Ho for
helpful discussions. This work was
completed during a visit to National Taiwan University, we thank
P-M. Ho, W-Y. P. Hwang and Y-C. Kao for hospitality.

\vfill
\eject

\listrefs
\end